\PassOptionsToPackage{unicode}{hyperref}
\PassOptionsToPackage{hyphens}{url}
\documentclass[
]{article}
\usepackage{amsmath,amssymb}
\usepackage{iftex}
\ifPDFTeX
  \usepackage[T1]{fontenc}
  \usepackage[utf8]{inputenc}
  \usepackage{textcomp} % provide euro and other symbols
\else % if luatex or xetex
  \usepackage{unicode-math}
  \defaultfontfeatures{Scale=MatchLowercase}
  \defaultfontfeatures[\rmfamily]{Ligatures=TeX,Scale=1}
\fi
\IfFileExists{upquote.sty}{\usepackage{upquote}}{}
\IfFileExists{microtype.sty}{% use microtype if available
  \usepackage[]{microtype}
  \UseMicrotypeSet[protrusion]{basicmath} % disable protrusion for tt fonts
}{}
\makeatletter
\@ifundefined{KOMAClassName}{% if non-KOMA class
  \IfFileExists{parskip.sty}{%
    \usepackage{parskip}
  }{% else
    \setlength{\parindent}{0pt}
    \setlength{\parskip}{6pt plus 2pt minus 1pt}}
}{% if KOMA class
  \KOMAoptions{parskip=half}}
\makeatother
\usepackage{xcolor}
\usepackage[margin=1in]{geometry}
\usepackage{longtable,booktabs,array}
\usepackage{calc} % for calculating minipage widths
\usepackage{etoolbox}
\makeatletter
\patchcmd\longtable{\par}{\if@noskipsec\mbox{}\fi\par}{}{}
\makeatother
\IfFileExists{footnotehyper.sty}{\usepackage{footnotehyper}}{\usepackage{footnote}}
\makesavenoteenv{longtable}
\usepackage{graphicx}
\makeatletter
\def\maxwidth{\ifdim\Gin@nat@width>\linewidth\linewidth\else\Gin@nat@width\fi}
\def\maxheight{\ifdim\Gin@nat@height>\textheight\textheight\else\Gin@nat@height\fi}
\makeatother
\setkeys{Gin}{width=\maxwidth,height=\maxheight,keepaspectratio}
\makeatletter
\def\fps@figure{htbp}
\makeatother
\usepackage[OT1,T1]{fontenc}
\usepackage[utf8]{inputenc}
\usepackage[scale=1.05]{crimson}
\usepackage[scale=1.05]{biolinum}
\usepackage[scale=0.95,varl]{inconsolata}

\allowdisplaybreaks
\newcommand\numberthis{\addtocounter{equation}{1}\tag{\theequation}}

\usepackage[format=plain,
  labelfont={bf,sf,small,singlespacing},
  textfont={sf,small,singlespacing},
  justification=justified,
  margin=0.25in]{caption}

\usepackage[compact]{titlesec}
\usepackage{sectsty}

\renewenvironment{abstract}{
  \centerline
  {\large\sffamily\bfseries Abstract}
  \begin{quote}\small
}{
  \end{quote}
}

\allsectionsfont{\sffamily}

\usepackage{hyperref}
\ifLuaTeX
  \usepackage{selnolig}  % disable illegal ligatures
\fi
\usepackage[]{natbib}
\IfFileExists{bookmark.sty}{\usepackage{bookmark}}{\usepackage{hyperref}}
\IfFileExists{xurl.sty}{\usepackage{xurl}}{} % add URL line breaks if available
\hypersetup{
  pdfauthor={Tyler Simko; Shiro Kuriwaki; Christopher T. Kenny ; Cory McCartan ; Kosuke Imai},
  pdfkeywords={Congressional elections, partisan bias, redistricting, representation, Monte Carlo simulation},
  hidelinks,
  pdfcreator={LaTeX via pandoc}}

\title{\textsf{\textbf{
Widespread Partisan Gerrymandering Mostly Cancels Nationally,
but Reduces Electoral Competition}}}
\author{Christopher T. Kenny\footnote{%
  Department of Government, Harvard University}
$^\ddagger$ \and Cory McCartan\footnote{%
Department of Statistics, Harvard University}
$^\ddagger$ \and Tyler Simko$^*$\footnote{%
C.T.K., C.M., and T.S. contributed equally to this work.}
\and Shiro Kuriwaki\footnote{Department of Political Science, Yale University} \and Kosuke Imai$^{*\dagger}$\thanks{%
To whom correspondence should be addressed. E-mail: \texttt{\url{imai@harvard.edu}}.
C.T.K. has served as a paid expert for the Maryland Redistricting Commission. K.I. has served as a paid expert witness in the court cases related to the 2020 Congressional redistricting in Alabama, Kentucky, Ohio, and South Carolina.}}
\date{\today}

\begin{document}
\maketitle
\begin{abstract}
Congressional district lines in many U.S. states are drawn by partisan actors, raising concerns about gerrymandering. To separate the partisan effects of redistricting from the effects of other factors including geography and redistricting rules, we compare possible party compositions of the U.S. House under the enacted plan to those under a set of alternative simulated plans that serve as a non-partisan baseline. We find that partisan gerrymandering is widespread in the 2020 redistricting cycle, but most of the electoral bias it creates cancels at the national level, giving Republicans two additional seats on average. Geography and redistricting rules separately contribute a moderate pro-Republican bias. Finally, we find that partisan gerrymandering reduces electoral competition and makes the partisan composition of the U.S. House less responsive to shifts in the national vote.
\end{abstract}

\textbf{\textit{Keywords}}\quad
Congressional elections \textbullet~
partisan bias \textbullet~
redistricting \textbullet~
representation \textbullet~
Monte Carlo

\urlstyle{tt}

\hypertarget{introduction}{%
\section{Introduction}\label{introduction}}

A party that draws its own districts is likely to engage in partisan \emph{gerrymandering}---the drawing of district lines by a partisan actor to eliminate districts favorable to the opposing party and insulating their own incumbents from tough elections.
Based on the 2020 decennial Census, every U.S. state has recently redrawn their congressional districts lines, which shape the control of the House of Representatives for the next decade.
Scholars debate the extent to which parties have gerrymandered district lines to their advantage \citep{chen_cottrell_2016,wktcho_2016, rodden2019}, and whether courts and reforms, which move map-drawing powers from legislatures to commissions, can prevent such gerrymandering \citep{cain_2012,edwards_2017,henderson_2018,zhang_2021}.

Unfortunately, neither identifying gerrymanders nor quantifying their biases in electoral outcomes are straightforward.
Americans are geographically sorted and segregated along both partisan and racial lines \citep{schelling_1971,massey_denton_1988,trounstine_2018,brown_enos_2021,brown_2021,mettler_2022}.
Congressional elections, however, occur in winner-take-all, single-member districts.
When this sorting is combined with districts, Democratic votes turn into seats less efficiently than Republican votes \citep{chen2013,rodden2019}.
As such, comparing a party's share of seats to its vote share within a state cannot establish if a districting plan, rather than political geography, systematically advantages one party over the other.

Comparing districting plans across states or time periods is equally fraught.
The geographic clustering of voters differs both across states and over time.
Cross-state comparisons further mask differences in the rules for drawing districts.
For example, some state laws limit the set of possible plans to those which preserve counties, encourage partisan competition, or are more geographically compact.
Federal laws, such as the Voting Rights Act of 1965 (VRA), also constrain the set of possible plans that a state may enact without facing litigation.

We quantify the partisan effects of redistricting separately from other sources of bias such as political geography and redistricting rules.
We achieve this by comparing potential electoral outcomes under enacted district plans to those under a set of alternative plans that are created by simulation.
Simulation techniques have recently been adopted widely by scholars, courts, and redistricting practitioners (e.g.~\citep{chen_2021,kenny_2021,best_2021}, \emph{Rucho v. Common Cause} (2019), \emph{Harper v. Hall} (2022), or \emph{League of Women Voters of Ohio v. Ohio Redistricting Commission} (2022)).
These simulations are drawn using the same geographic units and state-specific redistricting requirements as the enacted plan.
Therefore, any differences in the partisan outcomes between the enacted plan and the simulated, non-partisan baseline demonstrate the partisan effects of redistricting.\footnote{As noted at the beginning of this section, we call a redistricting plan a partisan gerrymander if partisan actors draw district boundaries to create an electoral advantage for their own party.
  According to this definition, plans drawn by non-partisan actors may have partisan effects but are not a partisan gerrymander.
  Moreover, in litigation, demonstrating both partisan effects (consequences) and partisan intent may be needed to establish that a particular plan is a gerrymander.
  We do not study the partisan \emph{intent} of map drawers.}

As explained in the Supplementary Information (SI), we improve methodologically upon a simulation approach used by other scholars to study redistricting at a national scale \citep{chen_cottrell_2016}.
Specifically, we use simulation methods that are designed to produce a representative sample from the relevant universe of plans \citep{fifield2020enum,smc}.
We also differ from \citep{warshaw_2022} in that we use the complete set of simulated plans from \citep{50stateSimulations} to study state-by-state and district-by-district partisan effects.

We find that the new 2022 congressional map is biased in favor of the Republican party, but this bias is similar in magnitude to the expected structural and geographic bias predicted by the simulated plans.
To win a majority in the US House of Representatives under the enacted plan, Democrats need more than 51.1\% of the national two-party popular vote, just 0.14 percentage points more than under the non-partisan baseline.
While both parties engage in partisan gerrymandering in many states, the resulting bias mostly cancels at the national level, giving Republicans two additional seats.
The remaining Republican advantage may be explained in part by other features like the geographic distribution of voters \citep{chen2013, rodden2019} and redistricting rules.

Our state-by-state analysis shows that Republicans made large gains in states like Texas, Florida, and Ohio, by packing urban Democrats.
In contrast, Democrats made many smaller gains in states like Illinois, North Carolina, Pennsylvania, and Michigan.\footnote{As described in more detail below and in Materials and Methods, our simulations incorporate federal and state-specific requirements for map drawing \citep{50stateSimulations}. For example, these requirements include particular goals (such as Colorado, which requires map-drawers to maximize the number of politically competitive districts) and particular measures (such as Iowa, which has legal requirements on how to measure compactness) when specified. Michigan requires partisan fairness as part of redistricting rules. We did not directly incorporate this criteria for two reasons. First, it is not clear which partisan fairness metric should be used. Second, the non-partisan nature of our simulated redistricting plans can be considered accounting for partisan fairness at least to some extent.}
As a result of state-level gerrymanders by both parties, the overall competitiveness and responsiveness of the House are lower than our non-partisan baseline.
We find that an additional percentage point increase in national popular vote nets each party only 7.8 seats, on average, versus 9.2 under the non-partisan baseline.

\hypertarget{partisan-gerrymandering-is-widespread-but-bias-mostly-cancels-at-the-national-level}{%
\section{Partisan gerrymandering is widespread, but bias mostly cancels at the national level}\label{partisan-gerrymandering-is-widespread-but-bias-mostly-cancels-at-the-national-level}}

For each state, we use an election model (detailed in the Materials and Methods section and SI), building on data from \citep{vest_2016,vest_2020,medsl-pres-county}, to compute the range of House seats the two parties are expected to win under each of the simulated, non-partisan plans from \citep{50stateSimulations}.\footnote{These estimates are designed for cross-state comparability and do not align exactly with the corresponding estimates included in \citep{50stateSimulations}, which incorporates state offices.}
These simulated plans incorporate each state's specific requirements for map-drawing, along with federal requirements, to ensure that the sample of simulated plans are representative of the space of legal plans \citep{50stateSimulations, smc}.

We then compare, for each state, the predicted electoral outcomes based on simulated districts with those under the enacted plan.
The resulting differences in electoral outcomes can be interpreted as evidence of partisan effect (beyond political geography and redistricting rules) because the simulated plans are generated under constraints which correspond to the redistricting requirements of each state.
This comparison to a complete, national set of 2020 non-partisan baseline plans from \citep{50stateSimulations} differentiates our findings from existing estimates of national biases in House elections \citep{warshaw_2022, economist2022, rakich2022} (see Section \ref{related-simulation-studies} of the SI for detailed comparison with existing simulation studies).
To be clear, while we can estimate differences in partisan outcomes, we cannot necessarily identify the intent of map-drawers, which may include goals beyond packing opposite party voters, such as protecting particular incumbents or suppressing the power of minority voters.

Furthermore, if a state's redistricting rules themselves impart a partisan bias, our analysis would not detect that bias, as the current legal requirements are used to construct the baseline.
Similarly, compliance with the VRA can have impacts on individual states' partisan compositions \citep{chen_2021}.
As such, our estimates are best interpreted as a measure of partisan bias given current political geography and state's redistricting rules, either of which could change in the future.

Our simulation methodology, however, can examine the potential partisan impact of these redistricting rules.
In Section \ref{impacts-of-redistricting-rules} of the SI, we illustrate this point by removing some redistricting rules and examining how doing so can alter our empirical findings.

\begin{figure}
\centering
\includegraphics[width=4.0in]{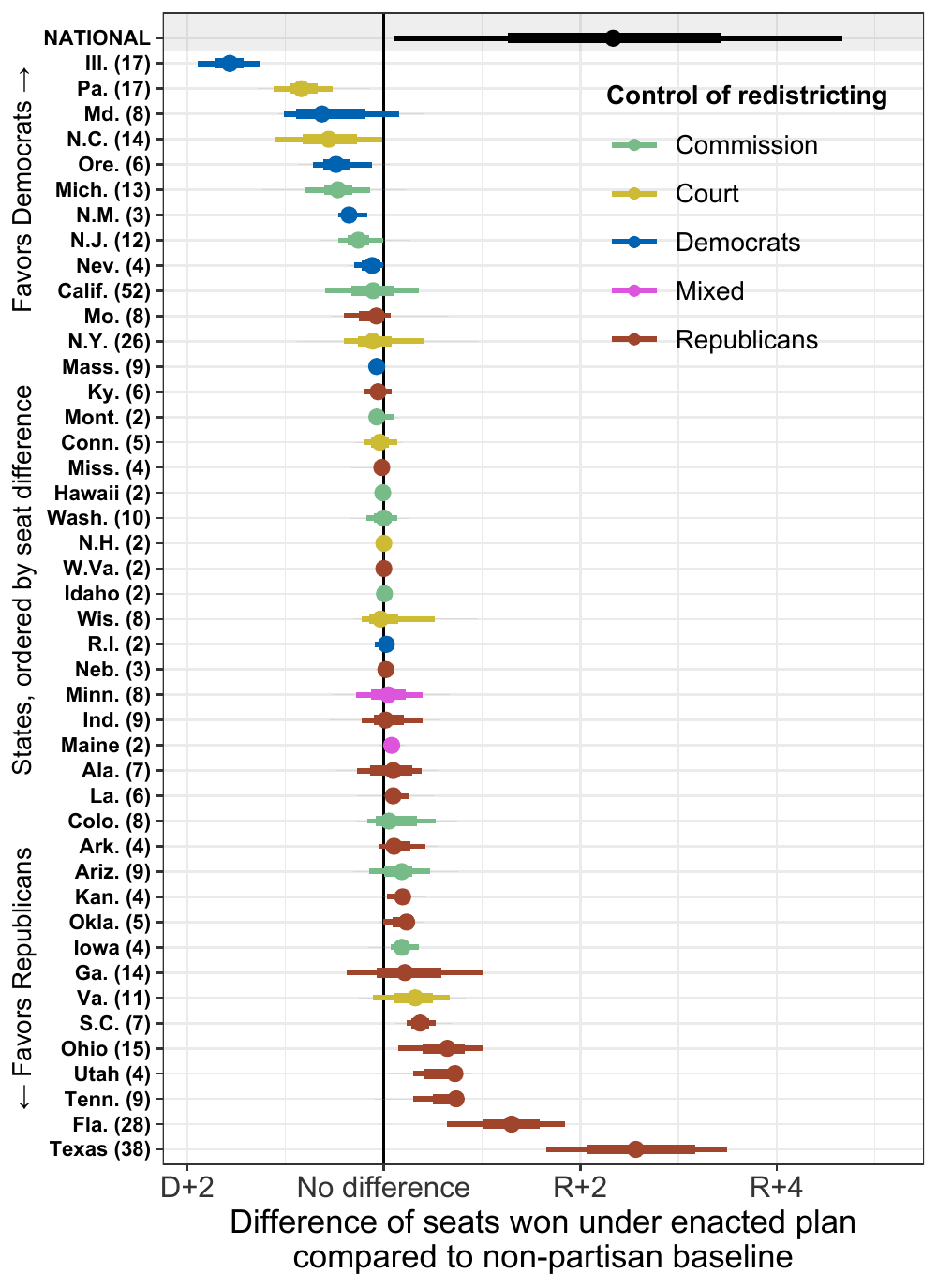}
\caption{Estimated state and national level partisan bias. Estimates show the expected Republican seats under each non-partisan baseline plan subtracted from the expected Republican seats under the enacted plan. Bars to the right indicate the enacted plan gives more seats to Republicans, while to the left indicates more seats for Democrats. The vertical black line indicates no partisan bias. The thick and thin bars show 66\% and 95\% intervals of the simulated plans, respectively, after averaging over uncertainty in future electoral swings at the national and district level. The point indicates the median. States are colored by the actors that controlled the map drawing in in each enacted plan, and are ordered vertically from the largest Democratic-favoring state to the largest Republican-favoring state. The number of seats in each state is shown in parentheses. In Section \ref{normalized-partisan-effect-plots} of the SI, we rescale the effects as a share of the state's seats and as $z$-scores.}
\label{fig:state-seats}
\end{figure}

Fig. \ref{fig:state-seats} shows clear evidence of widespread partisan gerrymandering.
Bars to the right of the vertical line indicate a Republican bias of the enacted plan, while bars to the left of the vertical line indicate a Democratic bias.
Whether the thin line crosses the zero line represents a statistical hypothesis test of the null of no difference at the 5\% level.
In 20 of the 44 states that redistricted, the interval excludes zero.
The intervals are not symmetric, as they are a function of each state's political geography, which motivates the use of sampled plans to construct the distributions.
We color each state by which party or institution drew that state's enacted plan and order the results vertically from the most Democratic-favoring state to the most Republican-favoring state.

Across all states, Fig. \ref{fig:state-seats} shows that partisan effects are expected to contribute to 8.6 Republican seats and 6.2 Democratic seats over a non-partisan baseline.
This nets out to a Republican advantage worth around 2.3 congressional seats.
Thus, the partisan bias created by widespread gerrymandering mostly cancels at the national level, while leaving Republicans slightly advantaged.
This pattern of cancellation amounting to a net Republican advantage is also found in an analysis of the 2010 redistricting cycle \citep{chen_cottrell_2016}.

The SI presents the same partisan effects as the share of each state's seats (as well as z-scores, in Figure \ref{fig:state-norm}) and in terms of the efficiency gap (Figure \ref{fig:state-egap}).
These results are consistent with our findings in Figure \ref{fig:state-seats}, showing that the states with the largest partisan biases tend to be those where a single party controls the redistricting process.

Pro-Republican bias is found primarily in the states where the Republican party controls the redistricting process.
For example, compared to the average of the simulated maps, the plan that the Texas legislature enacted is expected to net two additional seats for Republicans.
Florida exhibits a similar, but slightly smaller bias, leading to a just under a two seat advantage in most samples.
Ohio, South Carolina, Utah, Tennessee, Iowa, Kansas, and Louisiana also show smaller but statistically significant differences from the samples in favor of Republicans.

The maps drawn by Democratic state legislatures tend to show evidence of pro-Democratic bias.
These include Illinois, corresponding to slightly less than two seats, followed by Maryland, Oregon, New Mexico, and Nevada.

Although a majority of biased maps are drawn by state legislatures, some maps drawn by courts and commissions also exhibit statistically significant partisan effects (e.g., Pennsylvania, North Carolina, and Michigan for pro-Democratic maps, and Iowa for pro-Republican maps).
We do not treat these maps as partisan gerrymanders because they are drawn by non-partisan actors.
That said, commission-drawn maps are not free of bias, as shown by pro-Democratic bias in Michigan and the pro-Republican biases found for Iowa.
It is also an open question whether courts should be considered partisan actors in states such as North Carolina, where justices are elected in partisan elections.

\hypertarget{geographic-patterns-of-partisan-gerrymandering}{%
\section{Geographic patterns of partisan gerrymandering}\label{geographic-patterns-of-partisan-gerrymandering}}

\begin{figure*}[t]
\centering
\includegraphics[width=6.5in]{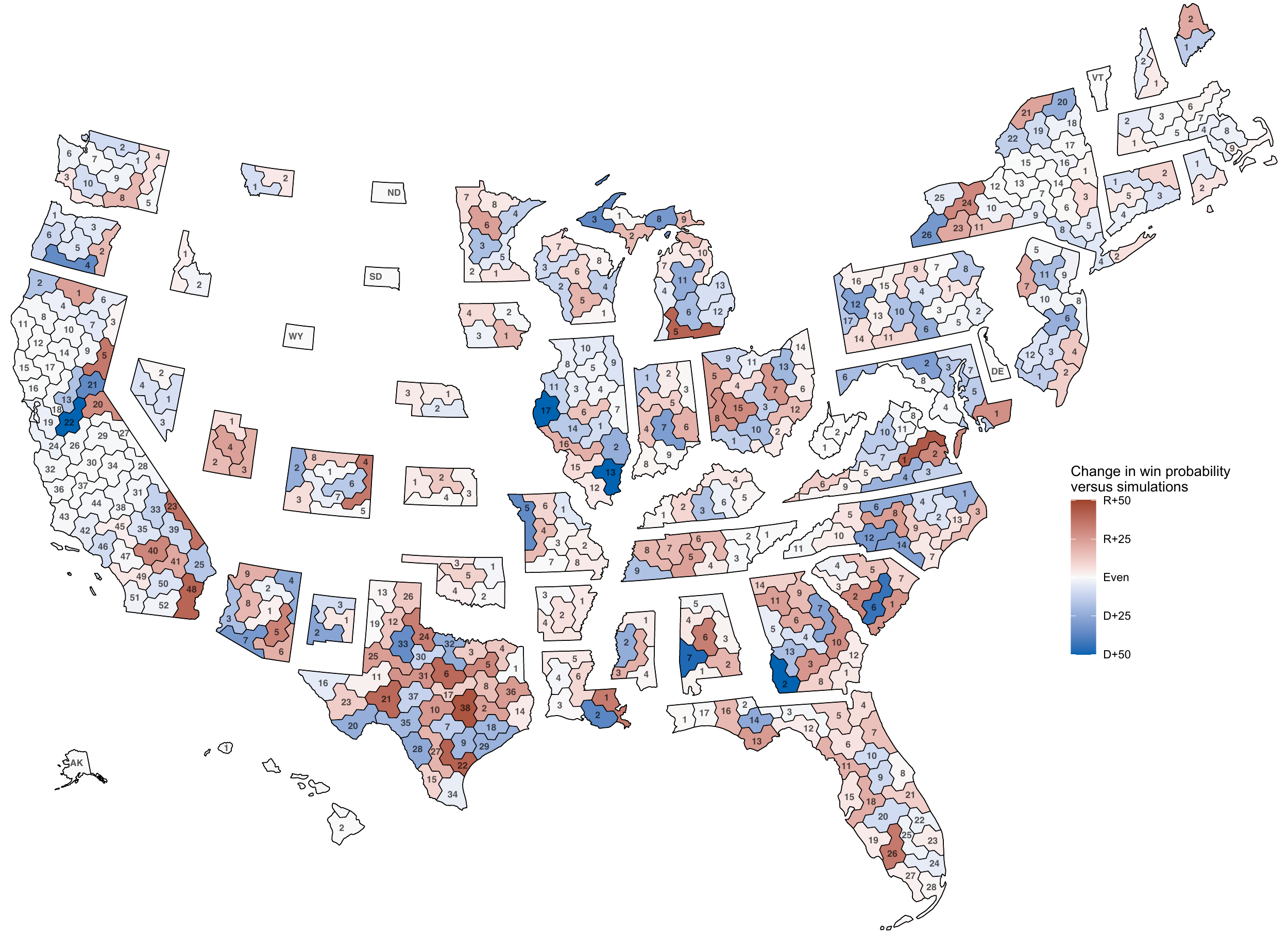}
\caption{Cartogram of new congressional districts, each
shaded by the difference between the probability of a party representing the district under the enacted plan and the voter-weighted average of the probability of a party representing the voters in that district under a non-partisan baseline.
See Section \ref{change-in-win-probability-map} of the SI for details on the computation of each probability.
Numbers represent the congressional district's number in the enacted plan.
Congressional districts are roughly equal in population so the cartograms are drawn to be roughly equal in size, while preserving the relative geographic locations of each district within the state. This cartogram style is inspired by the Daily Kos (http://dkel.ec/map).}
\label{fig:hex-manipulation}
\end{figure*}

How do states produce these partisan biases?
Our national-level results, demonstrating a slight bias in favor of the Republican party, align with those of existing estimates from \citep{warshaw_2022, economist2022, rakich2022}.\footnote{In a hypothetical tied national election, both \citep{economist2022} and \citep{rakich2022} predict Democrats winning 210 seats, with \citep{warshaw_2022} slightly more conservative at 205 seats.}
Our simulation approach, creating a set of 2020 counterfactual maps, allows us to further investigate the source of these biases at the state and district levels.
Fig. \ref{fig:hex-manipulation} disaggregates the estimates in Fig. \ref{fig:state-seats} to the district level, showing the partisan bias of districts in each state.

The map is colored by the difference in the probability of being won by Republican and Democratic candidates between the enacted map and the simulated plans.
Light colors indicate that the election outcome under the enacted plan is expected to be similar to the outcomes in the same geographic area across the non-partisan baseline plans, while darker colors indicate larger differences.
Districts in red indicate that an enacted district advantages Republicans compared to the simulated plans, while blue districts indicate a Democratic advantage.
As explained in the Materials and Methods section, this map builds upon recent work that focuses on precinct-level quantities, such as partisan dislocation \citep{deford2022partisan}, and is closely related to a general framework to evaluate districting plans at the individual voter level \citep{mccartan2022individual}.

We see that enacted plans in some states have only small differences from the simulated plans.
For example, Massachusetts and West Virginia each have only muted shades of red and blue---they have only small differences in each district.
But, other states have more substantial differences.
In North Carolina, where the state supreme court drew the map, its pro-Democratic bias stems in part from the drawing of districts 12 and 14 in the Charlotte area.
The enacted plan splits Charlotte through the middle, which creates two Democratic districts when combined with the suburbs.
In contrast, the simulated alternative plans typically draw a full district within Charlotte's enclosing county.

In Texas, the enacted plan strongly favors Republicans, as seen in Fig. \ref{fig:state-seats}.
The Texas legislature made two districts in the Houston area, 22 and 38, far safer for Republicans than expected.
This corresponds to the packing of urban Democratic voters in districts 7, 9, 18, and 29.
Some of these districts are overwhelmingly composed of racial minorities, and the VRA does compel states to draw such districts in some circumstances.
However, the enacted plan appears to pack these districts with Democratic voters far beyond what may be needed to ensure that the district usually elects minority voters' preferred candidates.
Similar approaches in Austin and Dallas areas cement the net bias towards Republicans.

\hypertarget{partisan-gerrymandering-reduces-electoral-competition-and-responsiveness}{%
\section{Partisan gerrymandering reduces electoral competition and responsiveness}\label{partisan-gerrymandering-reduces-electoral-competition-and-responsiveness}}

We also analyze the impact of partisan effects on each party's ability to translate votes into seats under different electoral environments.
Widespread gerrymandering could limit the electoral power of voters in many affected districts, even if biases mostly cancel out between parties at the national level.
We first estimate a baseline partisanship for each precinct by averaging the 2016 and 2020 presidential elections.
We then tally the baseline within each of the enacted and simulated districts to obtain an estimate of district-level baseline partisanship.
Finally, we use these estimates to examine electoral competitiveness under the enacted and simulated plans.

\begin{figure}
\centering
\includegraphics[width=3.25in]{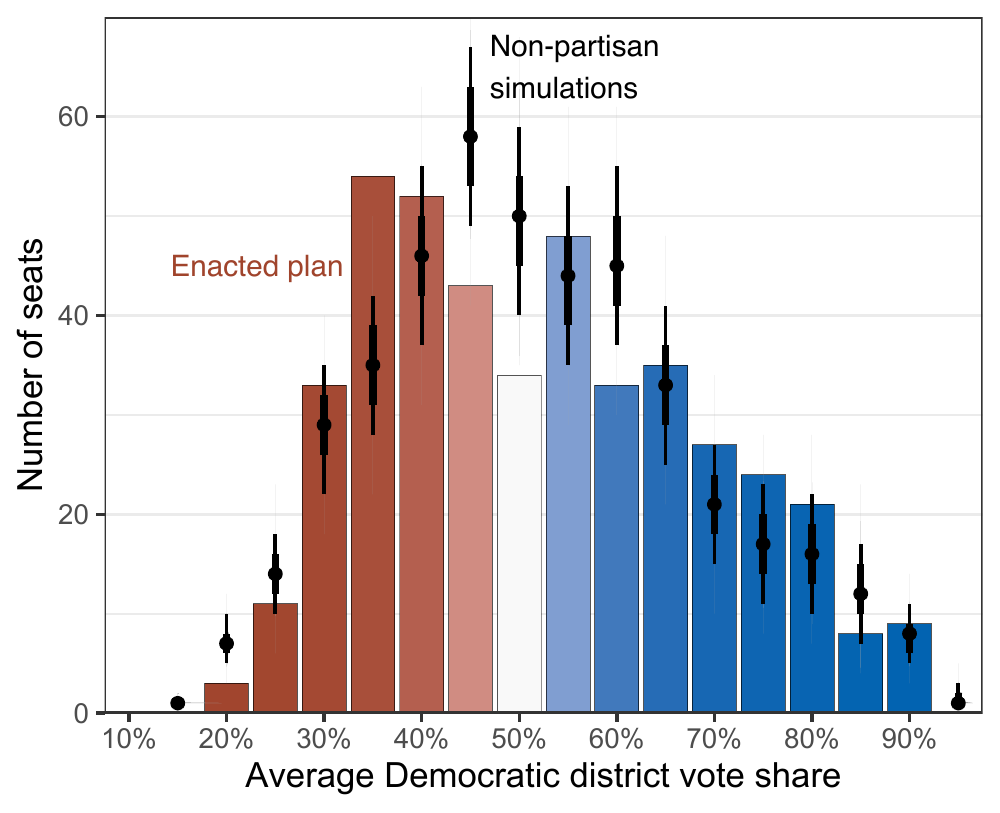}
\caption{The histogram shows the distribution of expected district voteshares under the enacted districts.
The overlaid 66\% and 95\% confidence intervals show the range of the same quantity under our non-partisan simulated baseline.}
\label{fig:compet}
\end{figure}

Fig. \ref{fig:compet} shows the distribution of district-level partisanship for the enacted and simulated maps.
The enacted plans across all states create significantly fewer highly competitive seats, where the baseline vote share is between 47.5\% and 52.5\%.
Only 34 out of 435 districts under the enacted districting plans fall into this category, compared to 50 in the non-partisan baseline.
There are fewer seats which lean Republican (52.5--57.5\% vote share) than expected, but \emph{more} safe Republican seats (62.5--67.5\% vote share) than expected.
This reflects the Republican gerrymandering strategy of shoring up Republican seats to insulate them from elections which swing towards Democrats.
In contrast, Democrats appear to have fewer moderately safe seats under the enacted plan than the simulated plans.

We can translate these baseline partisanship estimates into a national \emph{seats-votes curve}, which relates the national popular vote share of each party in House elections to the share of House seats they win under each redistricting plan \citep{tufte1973pb,katz2020}.
Such an analysis is useful since each future election may be subject to a different electoral environment (e.g., Republicans are generally expected to do well in a Democratic president's midterm election).
To do so, we use the election model detailed in the Materials and Methods section to calculate the expected number of seats each party would win under a range of national popular vote shares.

\begin{figure}[t]
\centering
\includegraphics[width=3.25in]{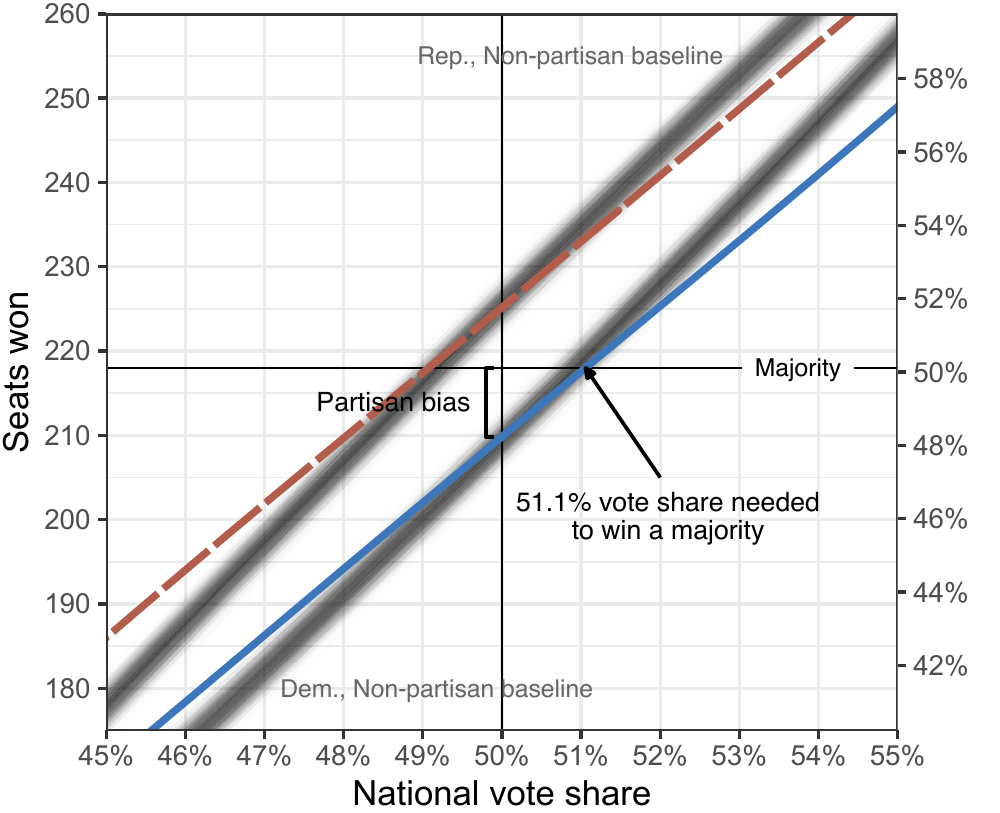}
\caption{Seats-votes curves that show how each party can translate a national popular vote into congressional seats.
The curve for Democratic seats and votes under the enacted plan is solid blue, while the curve for Republicans are dashed red.
Curves representing plans simulated under our non-partisan baseline are plotted as thin gray lines.
Partisan bias in this figure indicates the deviation from partisan symmetry evaluated at a voteshare of 0.5, following the definition of \citep{katz2020}.
}
\label{fig:seats-votes}
\end{figure}

Fig. \ref{fig:seats-votes} presents our estimate of the national seats-votes curve for each party under the enacted plans with the vote shares ranging from 45\% to 55\%.
The seats-votes curve under each of the 5,000 simulated non-partisan plans are shown in light gray.
The results show that in competitive elections, the seats-votes curve indicates a bias against Democrats.
If Democrats win 50\% of the popular vote, we estimate them to win around 210 out of 435 seats.
It is not until the Democratic party wins 51.1\% of the vote that they would receive half of the seats in the House.

This Democratic disadvantage, however, cannot be entirely attributed to partisan effects of redistricting.
Even if states did not draw a plan with partisan bias, the Democratic party would still be at a disadvantage.
Compared to the range of alternative seats-votes curves in gray, we see that if the national environment were to trend towards a 50\% national vote share between parties, the difference between the national bias and the baseline geographic and rule-based bias disappears.
Earlier work has shown that the geographic distribution of voters systematically benefits the Republican party \citep{chen2013, rodden2019}.
We find that this dynamic is likely to be still present in 2020 though the bias is due to the combination of political geography and redistricting rules.
On average, under our non-partisan baseline plans, Democrats must win 50.9\% of the popular vote in order to win half of the seats---very similar to the enacted plan.

Nevertheless, partisan bias in redistricting reduces electoral responsiveness.
Responsiveness describes the rate at which changes in the vote share result in changes in the seat share.
Notice that the enacted plan's seats-votes curve has a flatter slope than the simulated plans.
This indicates that the enacted plan is less \emph{responsive} to national swings in the vote in competitive elections.

Fig. \ref{fig:resp} in the SI shows this responsiveness, measured as the number of seats gained by Democrats for a one percentage point increase in vote share.
For typical vote shares (45--55\%), the enacted plan is around 16\% less responsive than a non-partisan baseline would predict.
That is, each additional percentage point of the national vote nets 7.8 seats, on average, under the enacted plan, compared to an expectation of 9.2 seats under the simulated plans.
This is a direct result of the smaller number of competitive seats in the enacted plan, as documented above.

\hypertarget{discussion}{%
\section{Discussion}\label{discussion}}

The boundaries of districts in the U.S. Congress are drawn by individual states, each with different political geography and redistricting rules.
Therefore, traditional approaches that seek to estimate the partisan effect of redistricting by leveraging information from other states and/or previous periods can potentially be misleading.
Simulation-based approaches address this issue, but prior attempts have neither fully accounted for the state-specific nature of these differences nor taken advantage of recent advances in redistricting simulation algorithms \citep{chen2013,chen_rodden_2015,chen_cottrell_2016}.
Improved data and methods used in our analysis can better characterize the magnitude of these biases for all fifty states, while incorporating the state-specific redistricting criteria \citep{50stateSimulations}.

We find that congressional districting plans from the 2020 redistricting process exhibit widespread partisan effects---often, the plans differ from the average sampled plan and, in many cases are statistical outliers, compared to the sampled plans.
We also find that partisan bias created by redistricting largely cancels out at the national level.
However, this does not mean that the partisan gerrymandering --- and the political contestation over redistricting --- was inconsequential.
Many states have enacted districting plans with partisan biases that decrease electoral competitiveness and responsiveness, limiting the voter's ability to hold politicians accountable.

The simulated baseline plans allow us to show that geographic and institutional features disadvantage Democrats, even absent partisan gerrymandering.
In fact, we show that these structural factors dominate the final Democratic disadvantage in the enacted plans.
Finally, our analysis finds partisan effects given the current geographic and institutional composition of the House. Map-drawers may also gain partisan advantage through manipulations of the redistricting process itself, as different rules may have biased partisan implications.
These findings illuminate the inherent difficulty of producing a fair national House map when it is drawn piecemeal by fifty autonomous political units.
More research is needed to further advance our understanding of the crucial relationship between votes and seats that structures our democracy.

\hypertarget{materials-and-methods}{%
\section*{Materials and Methods}\label{materials-and-methods}}
\addcontentsline{toc}{section}{Materials and Methods}

\hypertarget{data-availability}{%
\subsection*{Data Availability}\label{data-availability}}
\addcontentsline{toc}{subsection}{Data Availability}

All data and code necessary to replicate our analyses is available on the Harvard Dataverse at \url{https://doi.org/10.7910/DVN/JI1U8X}.

\hypertarget{statesimulations}{%
\subsection*{\texorpdfstring{\textsc{50stateSimulations}}{50stateSimulations}}\label{statesimulations}}
\addcontentsline{toc}{subsection}{\textsc{50stateSimulations}}

For our analyses, we rely on \textsc{50stateSimulations}, which contains 5,000 sampled redistricting plans and accompanying precinct-level election and demographic data for each state \citep{50stateSimulations}.
These data are publicly available on the Harvard Dataverse at \url{https://doi.org/10.7910/DVN/SLCD3E}.
The source code which generates the sampled plans is available on GitHub at \url{https://github.com/alarm-redist/fifty-states/}.
Election data were originally collected by the Voting and Election Science Team (\url{https://dataverse.harvard.edu/dataverse/electionscience}).

For the 44 states with more than one congressional district, \textsc{50stateSimulations} identifies the state's legal requirements, incorporates them into mathematical constraints, and samples plans from a target distribution that corresponds to those constraints using the open-source software \citep{redist}.
Requirements which involve a legal determination, such as compliance with the Voting Rights Act, are assumed met by the enacted plan.
The level of compliance is then matched for the simulated set.
Note that the legality of some of the enacted plans is still in dispute.
Our analysis does not attempt to address the legal issues raised in these court cases.
\textsc{50stateSimulations} further attempt to match non-partisan discretionary criteria, such as number of counties split, number of municipalities split, and compactness of districts.
The criteria incorporated, information as to what law or rule sets the requirement, and interpretation of the requirement for each state are available on the aforementioned Harvard Dataverse site, in each state's documentation file.
For the six states where there is only one congressional district (Alaska, Delaware, North Dakota, South Dakota, Vermont, and Wyoming), we supplement this with the only possible plan, repeated 5,000 times so that these districts are weighted equally in our analysis.

\hypertarget{electoral-modeling}{%
\subsection*{Electoral modeling}\label{electoral-modeling}}
\addcontentsline{toc}{subsection}{Electoral modeling}

To understand the electoral implications of 2020 redistricting plans in the upcoming decade, our findings rely on a statistical model of elections.
An election model uses observed past election results to quantify the uncertainty over future election results which take place under a different set of redistricting plans.

Our goal is to estimate the partisanship of each district in the enacted plan and also for alternative districts, with different geographic configurations, in each of the 5,000 simulated plans.
To do this, we use precinct-level election data and estimate the precinct-level baseline partisanship as an average of the 2016 and 2020 presidential elections.
We use previous presidential elections because the same candidate is on the ballot across the entire nation, unlike for Senate or House races.
This practice is also adopted by many elections analysts and is used in the Cook Partisan Voting Index.

Specifically, we take the mean of the Democratic two-party vote share in each precinct across the two elections, and separately take a geometric mean of the turnout across the elections (due to its skewed distribution), to produce a baseline number of Democratic and Republican votes for each precinct.
For example, the baseline Democratic vote count estimate for precinct \(j\), denoted by \(\hat D_j\), can be written as
\begin{align*}
\hat D_j &= \frac{1}{2}\left(\frac{D_{16j}}{D_{16j} + R_{16j}} +
    \frac{D_{20j}}{D_{20j} + R_{20j}}\right) \\
&\qquad \times \sqrt{(D_{16j} + R_{16j})(D_{20j} + R_{20j})}
\end{align*}
where \(D_{tj}\), \(R_{tj}\) are the Democratic and Republican vote counts for the precinct in year \(t\).
A geometric mean for turnout values corresponds to the usual mean for log-turnout values.
In Kentucky, some detailed 2020 election data is missing, and so we impute it from county-level and past precinct-level results, as described in Section \ref{missing-election-data-imputation} of the SI.

We model the two-party Democratic vote share in each district \(y_{it}\) for an election \(t\) as
\begin{align*}
  \mathrm{logit}(y_{it}) &= \alpha_i + \beta_t + \varepsilon_{it}, \numberthis\label{eq:elmod} \\
  \beta_t &\mathrel{\stackrel{iid}{\sim}} \mathcal{N}(0, \sigma^2_\beta), \\
  \varepsilon_{it} &\mathrel{\stackrel{iid}{\sim}} \mathrm{t}_\nu(0, \sigma^2_\varepsilon),
\end{align*}
where \(\alpha_i\) is district's baseline partisanship, \(\beta_t\) is election-to-election national swing, and \(\varepsilon_{it}\) is district-election specific error.
This is the Stochastic Uniform Partisan Swing model of \citep{gelman1994unified}, but put onto a logit scale.

Estimating this model poses challenges because of data limitations.
In particular, each different simulated plan has its own set of \(\alpha_i\) which must be estimated.
However, since these plans are hypothetical, we have no data on elections conducted under these plans.
So we fix \(\alpha_i\) for each district in the enacted plan and the 5,000 simulated plans by simply aggregating our estimate of the precinct-level baseline vote counts for the district.
Specifically, for a given plan \(p\), we compute
\[\hat{\alpha}_{ip} = \frac{\sum_{j \in \mathcal{J}_{ip}}{\hat{D}_j}}{\sum_{j \in \mathcal{J}_{ip}} \hat{D}_{j}  + \sum_{j \in \mathcal{J}_{ip}} \hat{R}_j}\]
where \(\mathcal{J}_{ip}\) indicates the set of precincts that are assigned to district \(i\) in a redistricting plan \(p\).

Additionally, since we use the model to predict \emph{future} elections, we will not know \(\beta_t\) and \(\varepsilon_{it}\).
Instead, these are drawn from the normal distributions with the variance of \(\sigma^2_\beta\) and \(t\)-distribution with \(\nu\) degrees of freedom and scale \(\sigma^2_\varepsilon\), respectively, under this model.
This injects the appropriate amount of uncertainty about future national and district-specific election swings into our election predictions, which are then propagated to uncertainty in our topline estimates that are presented as figures in the main text.

Thus, to create election predictions, it remains to estimate \(\sigma^2_\beta\), \(\sigma^2_\varepsilon\), and \(\nu\); once these are estimated along with our baseline estimates \(\hat{\alpha}_{i}\), we can simulate hypothetical future election outcomes.
To estimate \(\sigma^2_\beta\), \(\sigma^2_\varepsilon\), and \(\nu\), we fit the model given in \eqref{eq:elmod} to historical House elections.
The data \citep{medsl-house} contains almost all House elections since 1976.
We study only the races contested by exactly one candidate from each party.

In fitting this model, we are constrained by the lack of historical presidential election data disaggregated to the congressional district level.
This means that we cannot create estimates of \(\hat{\alpha}_i\) in the manner described above that we use for our future predictions.
Instead, in the historical election model only, we fit \(\alpha_i\) as a random effect, which is specific to each district but constant across elections.
To account for redistricting, which changes the districts every decade, we estimate a separate \(\alpha_i\) for each district-decade combination (for example, WA--07 from 2012--2020 would receive a single random intercept) as a random effect.
None of the \(\alpha_i\) estimated as part of our historical House election model are used in the predictions of future elections.
The use of random effects here is only to properly allocate the total variability in election returns to three sources: district-specific, year-specific, and district-year-specific effects.
Only the estimates of \(\sigma^2_\beta\), \(\sigma^2_\varepsilon\), and \(\nu\) are used to produce the results in the main text.
This modeling choice, as well as the overall predictive performance of the model, are investigated further in Section \ref{electoral-model-validation} of the SI.

Section \ref{app:avg-seats} of the SI also provides computational details on how we use the fitted election model to estimate the average number of seats Democrats will win under a particular plan.

The SI also provides computational details on how we use the fitted election model to estimate the average number of seats Democrats will win under a particular plan.

\bibliography{references}

\appendix
\clearpage

\hypertarget{supplementary-information}{%
\section{Supplementary Information}\label{supplementary-information}}

\setcounter{figure}{0}
\renewcommand{\thefigure}{S\arabic{figure}}

\hypertarget{details-of-the-simulation-algorithm}{%
\subsection{Details of the simulation algorithm}\label{details-of-the-simulation-algorithm}}

Here we provide a brief overview of the simulation algorithm detailed in \citep{smc}.

In contrast to many existing simulation approaches, which use Markov Chain Monte Carlo (MCMC), the algorithm uses Sequential Monte Carlo (SMC).
MCMC algorithms operate sequentially, starting with a valid redistricting plan and making changes to a district or pair of districts one at a time.
The SMC algorithm draws many maps in parallel, which reduces the overall dependence between plans and increases the efficiency of the algorithm.

The algorithm starts with many copies of a blank map (i.e., one with no districts drawn) and draws a district at random on each map.
This drawing process is achieved using a spanning-tree proposal distribution, which is similar to the spanning-tree proposal distribution used by some of the existing MCMC algorithms \citep{carter2019,deford2019}.
Based on the district drawn and the constraints that the researcher has specified for the simulation, the SMC algorithm assigns each plan a weight before moving on to the next district.
For the next and subsequent districts, partially-drawn plans are sampled from the previous iteration using their weights.
A random new district is drawn on each map and checked for conformity with the population tolerance constraint.
If a map does not satisfy the population constraint, it is discarded.
This process is repeated until enough plans have been generated for this step.
Weights are then recalculated for all the plans.

Overall, the algorithm repeats the splitting procedure \(n-1\) times, where \(n\) is the number of districts needed.
In the end, the algorithm outputs a set of sampled plans and weights.

While the algorithm has theoretical convergence guarantees, it is important to check for convergence or lack thereof in practice.
As detailed in \citep{smc} and \citep{50stateSimulations}, we perform an extensive battery of diagnostic checks to ensure the quality and accuracy of the samples we produce with the algorithm.

\hypertarget{app:avg-seats}{%
\subsection{Computing expected Democratic seats from the election model}\label{app:avg-seats}}

In principle, it is simple to compute the average number of seats Democrats will win under a particular redistricting plan.
We can simulate new vote shares for each district from the fitted election model.
Then for each simulation draw, we can compute the number of seats the Democrats win.
Averaging this number across simulation draws yields an estimate of the average Democratic seats.

However, this approach introduces Monte Carlo error controlled by the number of draws from the election model.
When the number of Democratic seats must be estimated for 50 states and nationwide, not just for the enacted plans but also for 5,000 simulated plans, it can prove computationally costly to have more than a few hundred draws from the election model.
The Monte Carlo error will then aggregate as we sum across districts and states, and will generally not be negligible compared to the variance from the redistricting simulation.

To address this issue, we can exactly compute the expectation of the number of Democratic seats with Gauss-Hermite quadrature.
Specifically, in a state with \(n\) districts, and given estimates \(\hat{\boldsymbol{\alpha}}\) of the baseline partisanship of each district, we wish to estimate
\begin{align*}
\mathbb{E}&\left[\sum_{i=1}^n \mathbf{1}\{y_{it}>\tfrac{1}{2}\}
    \,\middle|\, \boldsymbol{\alpha}=\hat{\boldsymbol{\alpha}}\right] \\
&= \sum_{i=1}^n \mathbb{E}[\mathbb{P}(y_{it}>\tfrac{1}{2}\mid
    \alpha_i=\hat{\alpha}_i,\beta_t)\mid \alpha_i=\hat{\alpha}_i] \\
&= \sum_{i=1}^n \mathbb{E}[\hat F_\varepsilon(\alpha_i+\beta_t)\mid \alpha_i=\hat{\alpha}_i],
\end{align*}
where the second step follows from the law of iterated expectations, and in the last step \(\hat F_\varepsilon\) is the CDF of the error term in the fitted election model (which is a plug-in estimate from the historical House election model).
Since the final expectation conditions on \(\alpha_i\), it is only averaging over \(\beta_t\), which is drawn from a \(\mathcal{N}(0, \sigma_\beta^2)\) distribution.
Thus we can estimate \(\mathbb{E}[\hat F_\varepsilon(\alpha_i+\beta_t)\,|\, \alpha_i=\hat{\alpha}_i]\) to high accuracy using Gauss-Hermite quadrature \citep{golub1969gh}, and without resorting to random draws from the election model.
Specifically, we use an order-6 approximation, which simulation testing showed was more than enough for extremely high accuracy.

\hypertarget{missing-election-data-imputation}{%
\subsection{Missing election data imputation}\label{missing-election-data-imputation}}

For Kentucky, where we are missing 2020 precinct-level data, we imputed precinct-level results by shifting the precinct-level vote shares from the 2016 election on the logit scale by the amount that the enclosing county swung (see \citep{rosenman2022} for a principled justification of this practice).
County-level data is available for both elections from \citep{medsl-pres-county}.
Specifically, we imputed \[
    \widehat{\frac{D_{20j}}{D_{20j} + R_{20j}}}
    = \mathrm{logit}^{-1}\left(
    \log\frac{D_{16j}}{R_{16j}}
    -\log\frac{D_{16c}}{R_{16c}}
    +\log\frac{D_{20c}}{R_{20c}} \right)
\] where the subscript \(c\) indicates the corresponding county-level data (recall that the log of the ratio of each party's vote is the logit of the two-party vote share).
We also shift turnout on the log scale in a similar fashion.

\hypertarget{electoral-model-validation}{%
\subsection{Electoral model validation}\label{electoral-model-validation}}

\begin{figure}[t]
\centering
\includegraphics[width=\linewidth]{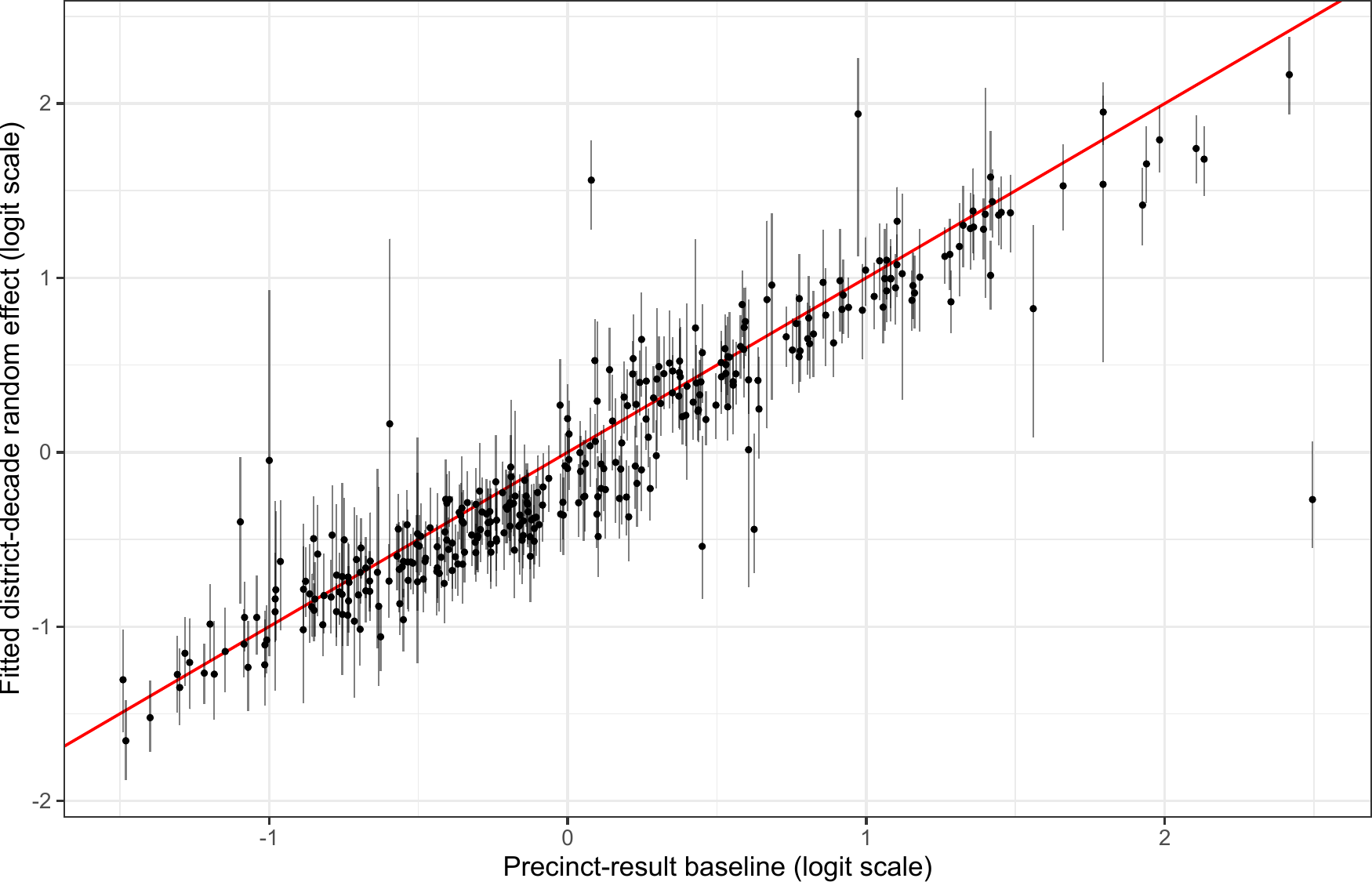}
\caption{Posterior point predictions and intervals for decade-district random effects, versus the 2016--2020 partisan baseline calculated from precinct-level presidential returns. The red line indicates perfect agreement.}
\label{fig:distr-ranef}
\end{figure}

It is important to validate the electoral model because fitting a random effect to past House elections is not the same as predicting a partisan baseline using past presidential results.
Figure \ref{fig:distr-ranef} shows the results of this validation.
We compare the fitted random effects for the 2010 decade-districts to the baseline that would be obtained by aggregating our precinct-level partisan baseline to 2010 district boundaries.
The agreement is generally excellent, despite the baseline only using information on two of the three presidential elections conducted during this period.
The few large deviations indicate particularly strong candidate or regional effects (like Democrats holding the Montana at-large district, for instance).

It is important to validate the electoral model because fitting a random effect to past House elections is not the same as predicting a partisan baseline using past presidential results.
Fig. \ref{fig:distr-ranef} shows the results of this validation.
We compare the fitted random effects for the 2010 decade-districts to the baseline that would be obtained by aggregating our precinct-level partisan baseline to 2010 district boundaries.
The agreement is generally excellent, despite the baseline only using information on two of the three presidential elections conducted during this period.
The few large deviations indicate particularly strong candidate or regional effects (like Democrats holding the Montana at-large district, for instance).

\begin{figure}[t]
\centering
\includegraphics[width=\linewidth]{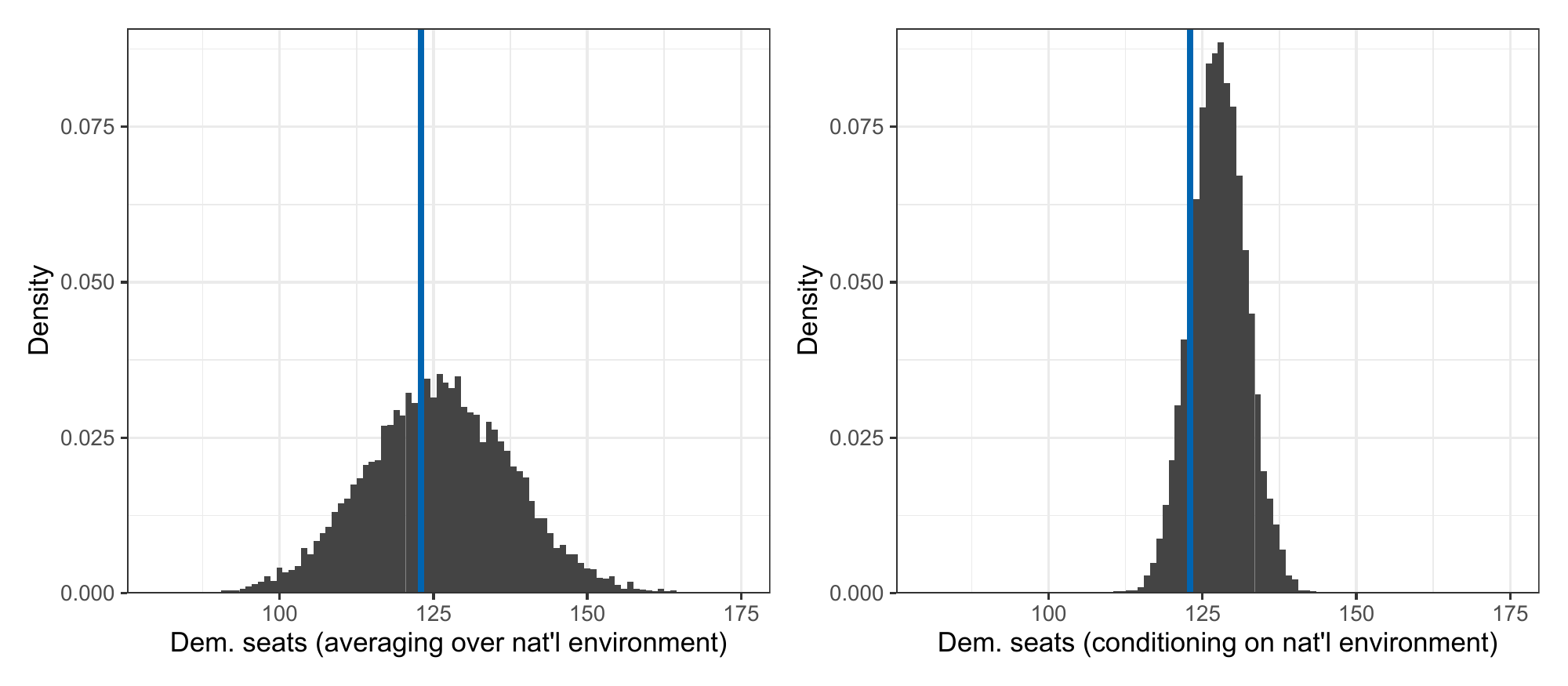}
\caption{Posterior predictive distribution and actual results for 255 2018 House elections in the historical dataset.
On the left, predictions are unconditional; on the right, they are conditional on the national environment.}
\label{fig:valid-18}
\end{figure}

It is difficult to further validate the predictive accuracy model, given the paucity of precinct-level election data with national coverage.
However, others have found the combination of heavy-tailed error and the Stochastic Uniform Partisan Swing model to produce well-calibrated predictions \citep{ebanks2022if}.
We also evaluate the predictions for 2018, both \emph{a priori} (i.e., averaging over possible national environments) and conditional on the actual shift from our 2016--2020 baseline.
Fig. \ref{fig:valid-18} shows these predictions, displayed as the number of seats Democrats are expected to win out of the 255 in our 2018 dataset.
For these predictions, we do not use the fitted district random effects (y-axis of Fig. \ref{fig:distr-ranef}), but rather the precinct-level partisan baseline (x-axis of Fig. \ref{fig:distr-ranef}).
The predictions are not fully out-of-sample, because 2018 data informs the overall estimates of \(\sigma_\beta\), \(\sigma_\varepsilon\), and \(\nu\).
The actual number of seats won by Democrats falls well within the range expected under the model.
Notice also the wider variation in the leftmost panel of Fig. \ref{fig:valid-18} due to the uncertainty in the national environment.

We fit the model with the \texttt{brms} R package \citep{brms}, with a weakly informative \(\mathrm{t}_3(0, 2.5)\) prior on the Intercept and all scale parameters, and a \(\mathrm{Gamma}(2, 0.1)\) prior on \(\nu\).
The model's estimates for \(\sigma_\beta\), \(\sigma_\varepsilon\), and \(\nu\) are summarized in Table \ref{tab:model-ests} below.
Estimates for the \(\beta_t\) are shown in Fig. \ref{fig:mod-yr}.
There is no evidence for heteroskedasticity corresponding to larger or smaller national swings over time.
This supports our use of historical data back to 1976.

\begin{table}[h]
\centering
\caption{Election model hyperparameter estimates based on a hierarchical model fit to historical data \label{tab:model-ests}}
\small

\begin{tabular}[t]{lrr}
\toprule
  & Estimate & Std. Error\\
\midrule
$\sigma_\varepsilon$ & 0.180 & 0.003\\
$\sigma_\beta$ & 0.117 & 0.021\\
$\nu$ & 2.896 & 0.128\\
\bottomrule
\end{tabular}

\end{table}
\begin{figure}[h]
\centering
\includegraphics[width=\linewidth]{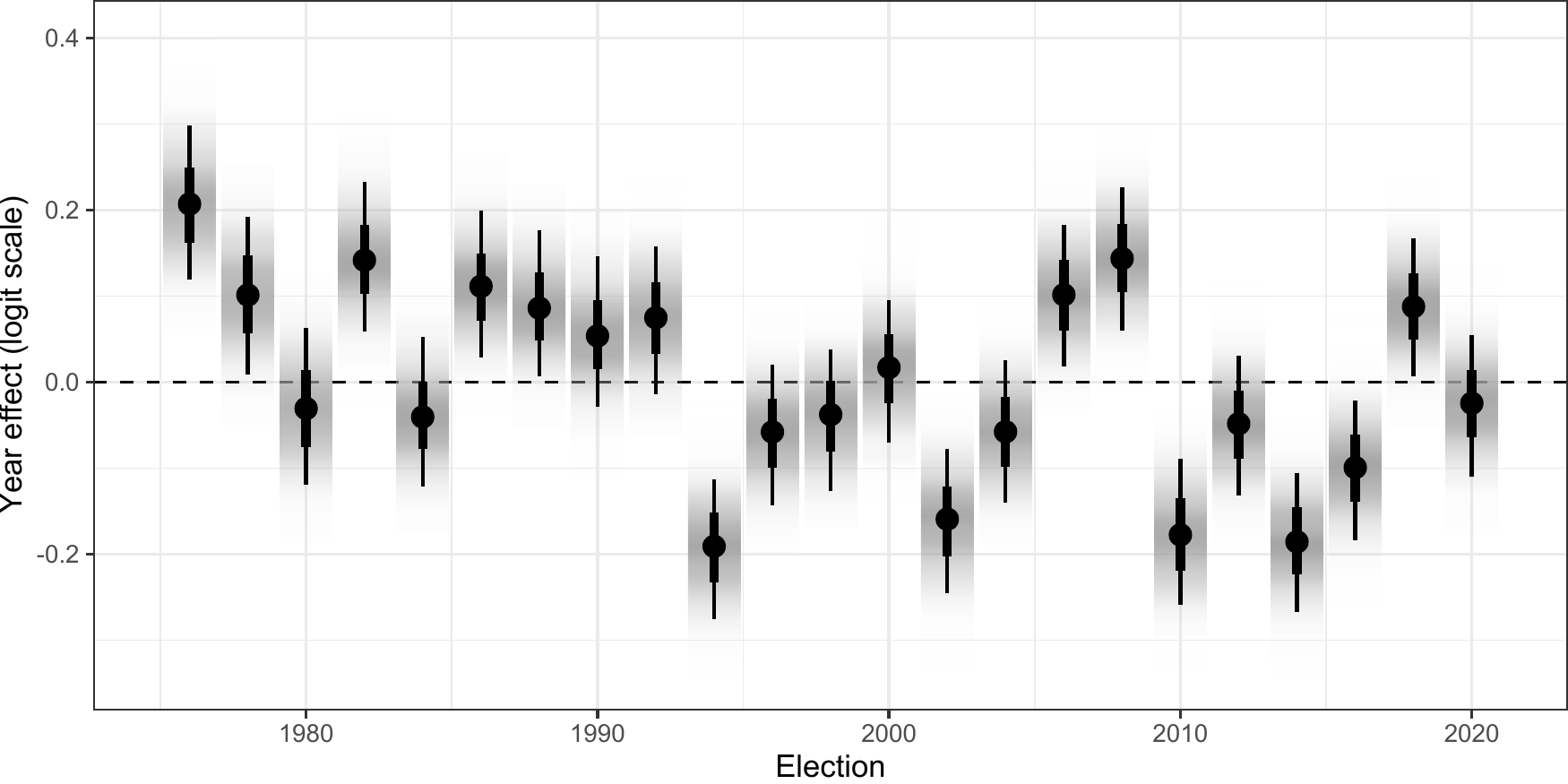}
\caption{Posterior distribution of $\beta_t$ for each election year, from a hierarchical model fit to historical data. The $\beta_t$ coefficients are on a logit scale, so a value of 0.1 corresponds roughly to a 2.5 percentage point vote share overperformance by Democrats.}
\label{fig:mod-yr}
\end{figure}

We use the posterior medians of these hyperparameters from the fitted model in our electoral predictions.
Combined with the baseline vote estimate \(\hat{\alpha}_{ip}\) for a specific plan \(p\), the model yields the posterior predictive distribution of the vote share \(y_{it}\) in district \(i\).
Averaging over this distribution gives us the probability that the district for any plan is won by a Republican.
This means that our point estimates of future election results are essentially equivalent to the Democratic performance in our past two Presidential elections, corresponding to a 51.5\% Democratic national vote.

As an example of the model's predictions, take GA-12, which covers eastern and southeastern Georgia.
GA-12 has a partisan baseline of 56.1\% Republican---the district's average vote share across the 2016 and 2020 presidential elections, under the new 2020 decade boundaries.
Averaging across all future hypothetical electoral environments, this translates to a 83\% chance that Republicans will represent the seat.
Conditioning on a particular national environment---say, a 5 percentage point shift towards the Democrats (a shift of 0.2 on the logit scale)---the model predicts a 59\% chance of a Republican win.

A key advantage of our modeling approach is that it propagates uncertainty across elections, in both national swing and district-level idiosyncrasies, into all of our paper's findings.
In particular, a district with a 50.01\% Democratic vote share at baseline will essentially count as half a seat for each party, since when averaging across simulations and electoral environments, half of the realizations will yield a Democratic victory and half will not.

\hypertarget{normalized-partisan-effect-plots}{%
\subsection{Normalized partisan effect plots}\label{normalized-partisan-effect-plots}}

\begin{figure}[t]
\centering
\includegraphics[width=\linewidth]{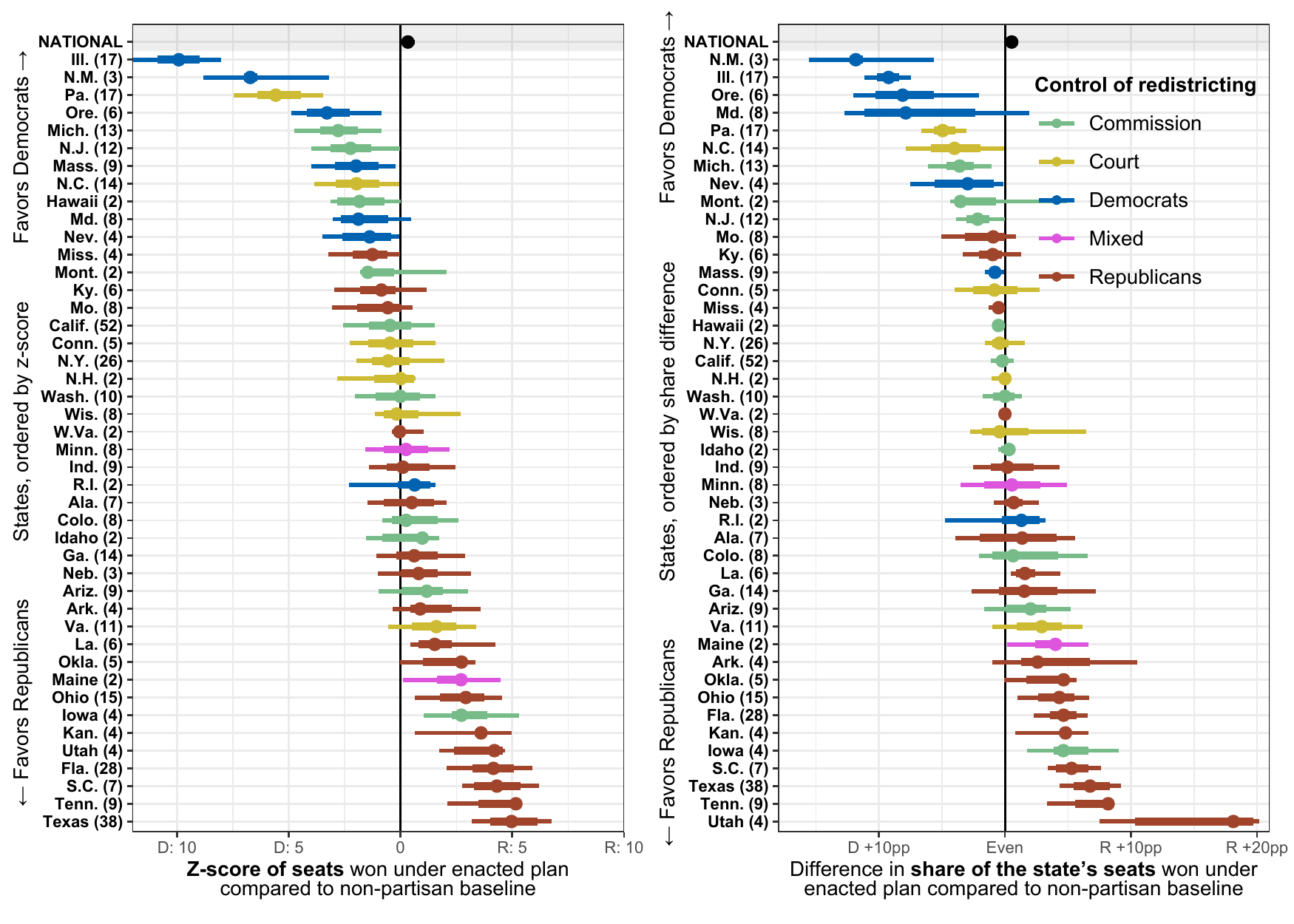}
\caption{Normalized estimates of state and national level partisan effects. Estimates show the expected Republican seats under each non-partisan baseline plan subtracted from the expected Republican seats under the enacted plan, normalized by the standard deviation of the simulated distribution (left) and by the total number of seats in the state (right). Bars to the right indicate the enacted plan gives more seats to Republicans, while to the left indicates more seats for Democrats. The vertical black line indicates no partisan effect. The thick and thin bars show 66\% and 95\% intervals of the simulated plans, respectively. States are colored by the actors that controlled the map drawing in in each enacted plan, and are ordered vertically from the largest Democratic-favoring state to the largest Republican-favoring state. The number of seats in each state is shown in parentheses.}
\label{fig:state-norm}
\end{figure}

Fig. \ref{fig:state-norm} re-expresses the data shown in Fig. \ref{fig:state-seats} through two different normalization methods.
The left plot normalizes the difference in seats between the enacted plan and simulations in each state by the standard deviation of the simulated distribution, producing a \(z\)-score.
More extreme values indicate more extreme partisan effects relative to the variation of the simulations.
The right plot normalizes the seat difference by the total number of districts in each state, producing a percentage point (pp) difference in seat shares.
This has the tendency of shrinking the apparent magnitude of partisan effects in larger states like Texas, and increasing the apparent magnitude in smaller states like Utah.

\hypertarget{efficiency-gap}{%
\subsection{Efficiency gap}\label{efficiency-gap}}

The efficiency gap measures the inter-party net percentage of votes that do not contribute to winning an election, including all votes for losing candidates and excess votes for winning candidates \citep{stephanopoulos2015}.
Unlike partisan bias, it does not require a seats-votes curve and thus can be calculated on a state-by-state basis \citep{king1987democratic,katz2020}.

\begin{figure}[t]
\centering
\includegraphics[width=0.5\linewidth]{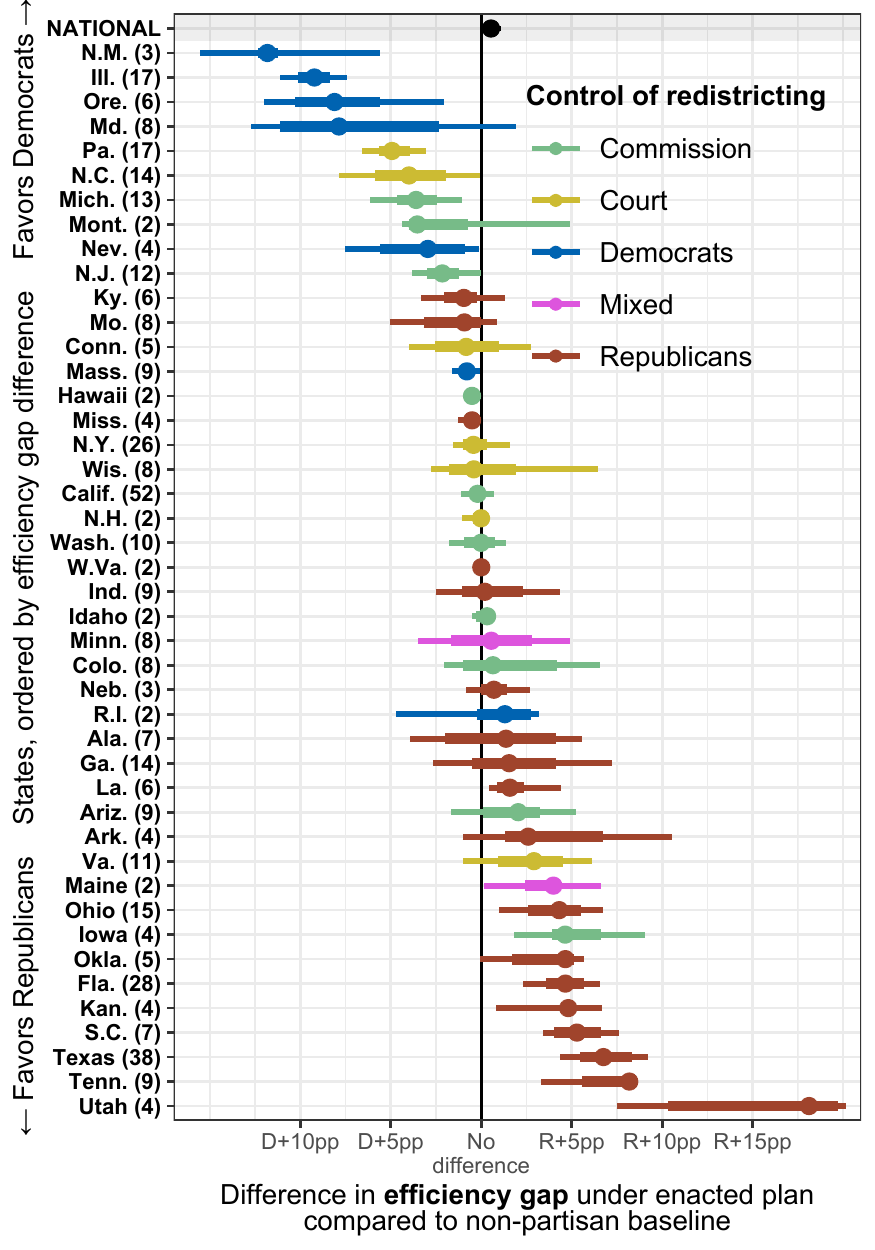}
\caption{Estimates of national and state level partisan effects measured in terms of efficiency gap. Estimates show the difference in efficiency gap under each non-partisan baseline plan subtracted from the expected efficiency gap of the enacted plan. Bars to the right indicate the enacted plan is more favorable to Republicans than expected, while bars to the left indicate the enacted plan is more favorable to Democrats than expected. The vertical black line indicates that the enacted plan aligns with the median simulated plan for the efficiency gap. The thick and thin bars show 66\% and 95\% intervals of the simulated plans, respectively. The point indicates the median. States are colored by the actors that controlled the map drawing in in each enacted plan, and are ordered vertically from the largest Democratic-favoring state to the largest Republican-favoring state. The number of seats in each state is shown in parentheses.}
\label{fig:state-egap}
\end{figure}

Fig. \ref{fig:state-egap} shows a measure of partisan effect in terms of the difference in the efficiency gap between the enacted plan and simulated non-partisan plans, instead of seats won.
To compute these differences, we follow the method from our main analysis.
For each national swing produced by the electoral model, we compute the efficiency gap for a plan.
We then average across those swings and examine the distribution of differences in averaged efficiency gaps across plans.
Although the efficiency gap is designed to have a ``fair'' value of 0, this value does not necessarily---and rarely does---align with the median of the non-partisan baseline simulations.
In addition, the confidence intervals for the efficiency gap are often asymmetric.

The findings are largely similar to those shown in Fig. \ref{fig:state-seats} as well as Fig. \ref{fig:state-norm}.
States such as Illinois and New Mexico, whose redistricting process is controlled by Democrats, exhibit a Democratic bias.
In contrast, Republican-controlled states including Texas and Utah show the partisan effect that favors Republicans.

\hypertarget{change-in-win-probability-map}{%
\subsection{Change in win probability map}\label{change-in-win-probability-map}}

\begin{figure}[t]
\centering
\includegraphics[width=\linewidth]{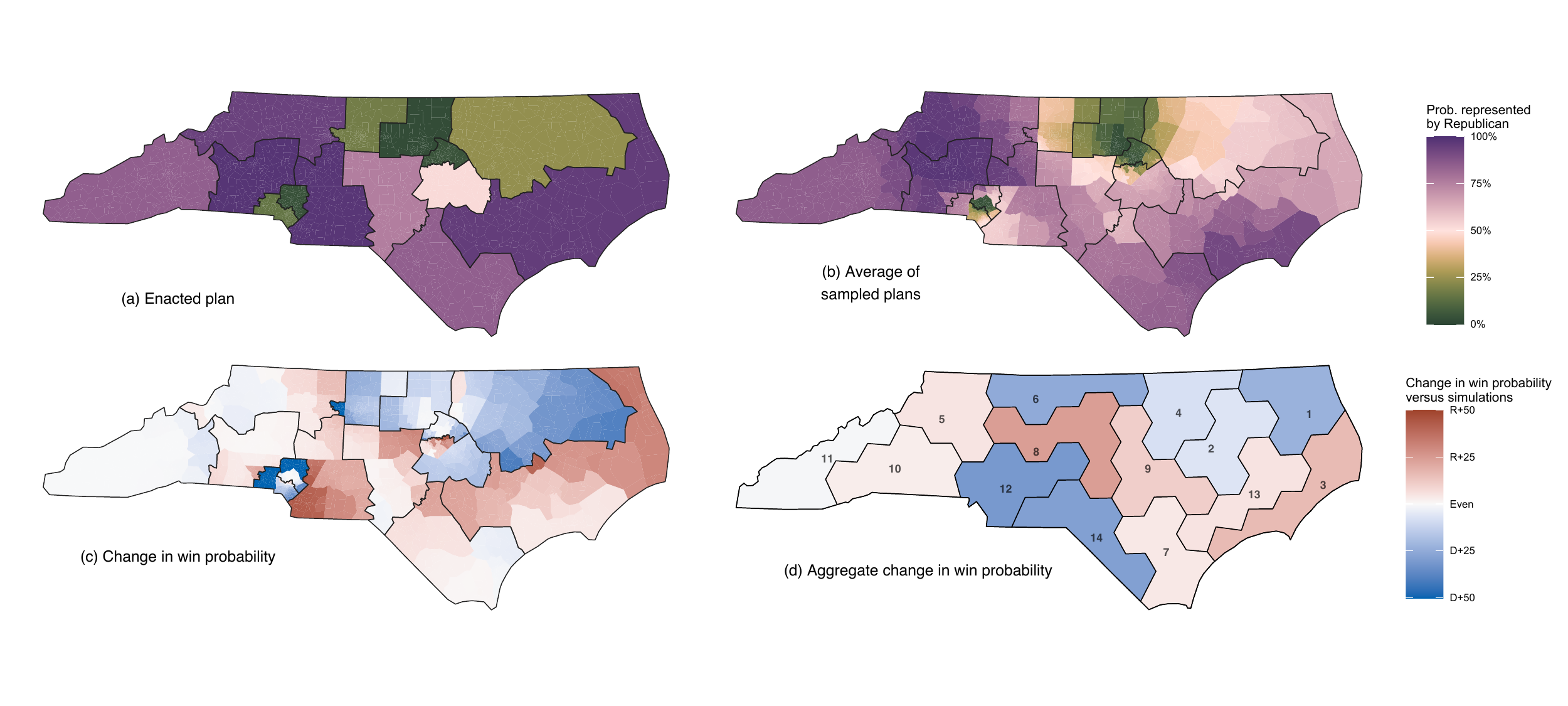}
\caption{Schematic illustrating the construction of Fig. \ref{fig:hex-manipulation}, using the state of North Carolina. Panel (a) represents the estimates from our election model under the enacted plan. Panel (b) represents estimates generated from the same model but under alternateive plans in our non-partisan baseline. These probabilities are estimated for each precinct.  The enacted district boudaries are overlaid for comparison.   Panel (c) is the difference of panels (a) and (b).  Finally, panel (d) averages the precinct-level differences in panel (c) across the voters in each enacted district and places the district in the roughly same area as the actual enacted district, using the same number of hexagons (reflecting the fact that each congressional district has a nearly equal population). These are shown in Fig. \ref{fig:hex-manipulation}.}
\label{fig:schematic}
\end{figure}

Each of our simulated non-partisan plans assigns individual voting precincts to a Congressional district.
For every one of the precincts, we first compute the probability that the encompassing district as a whole is expected to be won by a Republican candidate under the enacted plan (Fig. \ref{fig:schematic}(a)).
We repeat this for each of the simulated plans and average across the plans (Fig. \ref{fig:schematic}(b)).
We then take the difference of these two values to calculate the change in the probability that each precinct is represented by a Republican between the enacted and the simulated plans, on average (Fig. \ref{fig:schematic}(c)).
Finally, we take an average of these changes (weighted by the number of two-party voters in each precinct) within the enacted districts to form the change probability that a randomly sampled individual in an enacted district \(j\) is represented by a Republican.
This process is shown in Fig. \ref{fig:schematic}(d), and is what is plotted in Fig. \ref{fig:hex-manipulation}.

\hypertarget{seats-votes-curves}{%
\subsection{Seats-votes curves}\label{seats-votes-curves}}

The seats-votes curve is a commonly used visualization of district-based electoral systems.
Prior work has summarized past congressional elections \citep{kastellec_2008} and the electoral college with this approach \citep{cervas_grofman_2022}.
We make a new type of comparison, making thousands of seats-votes curves for each simulated plan to compare with the enacted national seats-votes curve.

Our seats-votes curve analysis examines how changing the national popular vote share using uniform swings affects the number of seats each party is predicted to win.
We estimate the curve from our election model as follows.
First, we fix a value of \(\beta_t\), the national swing.
Then, we estimate the number of seats won by each party using the election model using the baseline estimates \(\hat\alpha_i\) and the hyperparameter estimates, \(\hat\sigma_\varepsilon^2\) and \(\hat\nu\).
This produces one point on the seats-votes curve for each party.
We repeat this process across a wide range of \(\beta_t\), which traces out the entire seats-votes curve.

The seats-votes curve is closely tied to the theoretical framework of \emph{partisan symmetry} that is used as a measure of partisan fairness \citep{king1987democratic,katz2020}.
In this framework, if a districting plan is fair, parties are expected to receive the same number of seats for the same vote share.
That is, if party A receives \(v = 45\%\) of the vote share receives \(s\) seats, then when party B wins 45\% of the vote share (so party A wins \(1 - v = 55\%\)), it too should win \(s\) seats (leaving \(1 - s\) seats for party A).
That is, we say there is \emph{partisan symmetry} if \[S(v) = 1 - S(1 - v).\]

The deviation from symmetry is then measured for a given vote share as the number of seats that would need to be given from one party to another.
We call this deviation the \emph{partisan bias} of an electoral system \citep{king1987democratic}.
Factors like geographic bias, institutional procedures, and gerrymandering can result in biased seats-votes curves.
Note that in systems like the US House, there is no implicit guarantee of proportionality \citep{grofman_1983}.
The difference between a party's expected seat share from a given vote share is often referred to as ``partisan bias.''

Symmetry along the whole curve is rare.
In a two party system, relevant parties do not typically receive 5\% or 95\% of the votes, outside of rare, uncontested races.
Instead, we often zoom into the relevant vote shares near 50\%, where party competition occurs.
In Fig. \ref{fig:seats-votes}, we present the electorally relevant subsection.

\begin{figure}[t]
\centering
\includegraphics[width=0.9\linewidth]{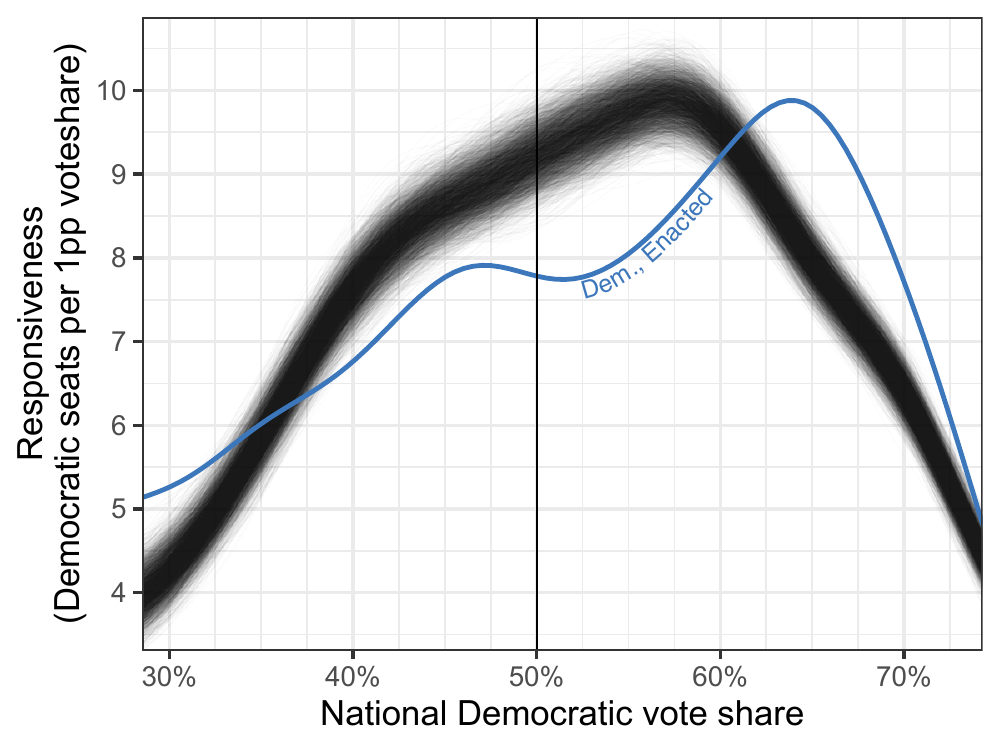}
\caption{Responsiveness of the electoral system under enacted and simulated plans. Responsiveness is measured as the change in seats for a one percentage point change in the national vote share.  It corresponds to the derivative of the seats-votes curves shown in Fig. \ref{fig:seats-votes}. For the realistic values of national Democratic vote share, responsiveness is much lower under the enacted plan (blue solid line) than under the simulated plans (grey solid lines).}
\label{fig:resp}
\end{figure}

A seats-votes curve encodes many different pieces of information.
While partisan bias describes the \emph{difference} between the seats-votes curves for each party at a particular vote share, the \emph{slope} of the seats-votes curve describes how responsive a system is.
Responsiveness captures for a percentage point increase in vote share, how many additional seats a party gains.
Our estimates of responsiveness for the national map and each of the simulated plans are shown in Fig. \ref{fig:resp}.
The figure shows that responsiveness is much lower under the enacted plan (blue solid line) than under the simulated plans (grey solid lines) in a realistic scenario where the national Democratic vote share is between 40\% and 60\%.

\hypertarget{related-simulation-studies}{%
\subsection{Related simulation studies}\label{related-simulation-studies}}

Our simulation approach improves upon that of \citep{chen_cottrell_2016}, which analyzed the previous 2010 redistricting cycle, in several ways.
First, we use a set of new samples of simulated, non-partisan plans from \citep{50stateSimulations}.
These simulated districting plans are tailored to each state after carefully constructing mathematical sampling constraints based on the legal requirements that each state imposes on its map drawers.
Second, the plans were simulated with a Sequential Monte Carlo (SMC) redistricting algorithm which has theoretical guarantees of representativeness \citep{smc}.
Third, the simulated plans are intended to be as compliant with the VRA as the enacted plans, though we do not attempt to resolve ongoing legal disputes about whether the enacted plans themselves are compliant with the VRA.
Finally, we use a Bayesian statistical model to generate a plausible range of future electoral outcomes primarily based on the two most recent presidential elections, rather than a more retrospective model from one election.
\citep{warshaw_2022} uses an earlier, though incomplete set of the simulated plans from \citep{50stateSimulations}, combined with data from PlanScore (planscore.org), which makes historical and cross-state comparisons to largely discuss state-by-state processes.
In contrast, we focus on the partisan biases at both national and state levels and isolate them from the structural biases, while incorporating simulation data from all states.

\hypertarget{impacts-of-redistricting-rules}{%
\subsection{Impacts of redistricting rules}\label{impacts-of-redistricting-rules}}

Simulated plans from \citep{50stateSimulations} are generated under the constraints consistent with the current federal and state redistricting rules.
This means that our estimates of partisan effects do not account for potential partisan bias due to these redistricting rules themselves.
We can assess the partisan effects of a particular rule of interest by adapting the simulation methodology used in our main analyses. Specifically, we run two sets of simulations for each state analysis---one of which includes a particular rule, and the other one does not---but are otherwise identical. Comparing these two sets of representative plans allows us to estimate the partisan effect of a rule itself.
In this section, we illustrate such an analysis through two examples: Ohio and Alabama.

\begin{figure}[t]
\centering
\includegraphics[width=3in]{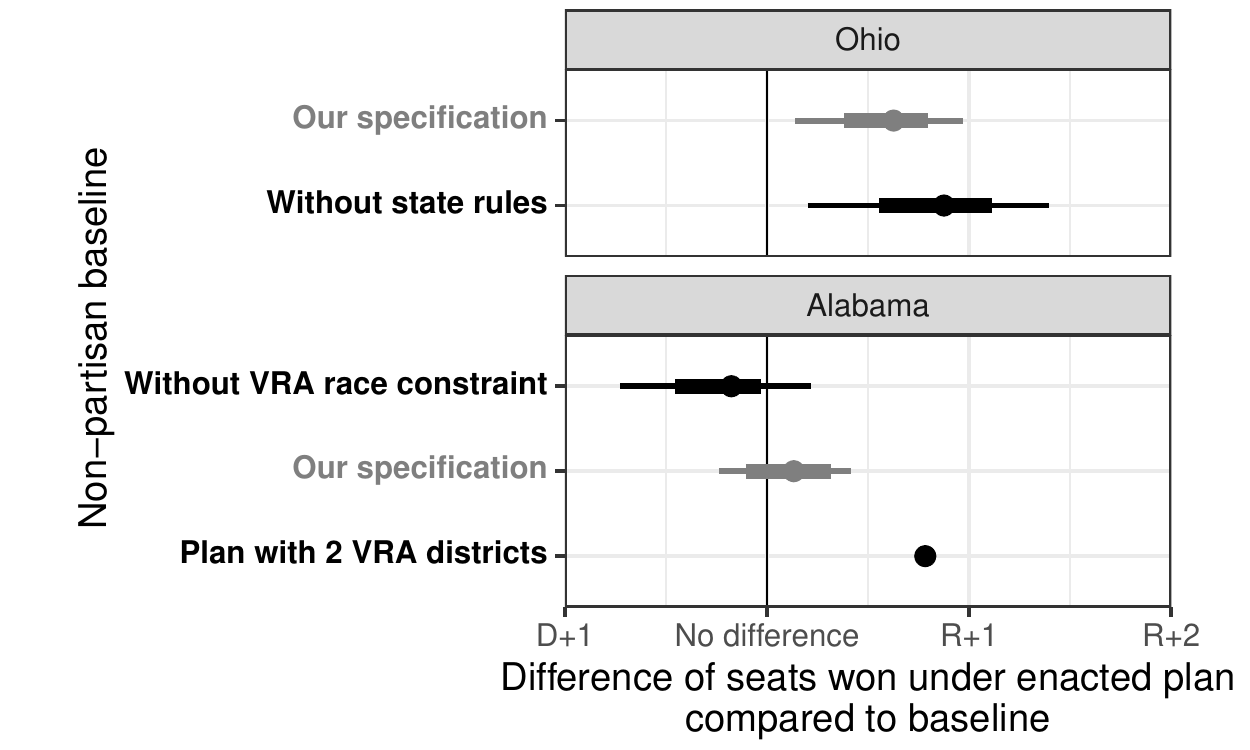}
\caption{Sensitivity of state-level analyses to difference in rules used. Each estimate shows the difference in seats won in the state given the enacted plan with a different choice of non-partisan baseline. The light gray specifications for Ohio and Alabama reproduce the results in Fig. \ref{fig:state-seats}. The alternative specifications in black use different constraints in the baseline simulation. Removing the rules increases our estimated Republican advantage of the Republican-drawn plan in Ohio and decreases it in Alabama, by modest amounts.}
\label{fig:endo}
\end{figure}

Ohio has historically been competitive statewide in some offices such as President, but the Republican party has maintained majorities in both of state legislative chambers and the Governor's office in 2010.
Therefore, Ohio is a state where partisan legislatures could bake in redistricting rules---such as those that require keeping large cities whole---that would bias our baseline simulations in the Republican party's favor.
In this case, keeping cities whole could prevent a more efficient distribution of Democratic-leaning urban voters across several districts \citep{rodden2019}.

To examine the sensitivity of our findings to these rules, we re-ran our simulation without constraints that limit the splitting of counties including Cincinnati, Cleveland, and Columbus, and used it as an alternative baseline.
We removed all constraints related to race and avoidance of administrative boundary splits, except for a constraint to limit the number of county splits to one less the number of total congressional districts (a constraint we use almost uniformly in other states).
Fig. \ref{fig:endo} shows that removing these rules increases the enacted plan's average Republican advantage by 0.25 congressional seats.
However, the variation in possible plans is relatively large compared to this difference in average.
In other words, on average, Ohio-specific rules may mask the extent of Republican gerrymandering, but the difference here is modest.

In Alabama, the key contention is over how the Voting Rights Act of 1965 should apply.
When Alabama drew the enacted plan, they drew one minority opportunity district, which is a Democratic district primarily controlled by Black voters.
Yet, in a court case currently at the US Supreme Court, \emph{Merrill v. Milligan}, the state has argued that they should only have to draw as many minority opportunity districts as would be generated by a race-blind analysis.
This could have partisan effects, due to the high correlation of race and party in Alabama, as it is possible that a race-blind process could draw no minority opportunity district.
Plaintiffs in the case instead argue that Alabama should have drawn two minority opportunity districts.

To examine the sensitivity of our analysis to these alternative contentions about the interpretation of the law, we created two more simulations as alternative baselines.
First, we create the race-blind specification which removes race entirely from the simulation parameters and uses standard parameters to preserve traditional redistricting criteria (``without VRA race constraint'' in Fig. \ref{fig:endo}).
Second, we create a specification ``with 2 VRA districts,'' which freezes the two minority opportunity districts from Plaintiff demonstration plan A in \emph{Merrill}.
This has an additional effect of freezing a third district, along the Gulf of Mexico, which is land-locked by the two frozen VRA districts.
Fig. \ref{fig:endo} shows that using a race-blind specification as a baseline would flip the bias of the enacted plan from favoring Republicans on average to favoring Democrats on average, though neither of these are statistically significant biases.
We note that this race-blind bias in favor of Democrats aligns with existing past simulation work that does not incorporate state-specific rules \citep{chen_cottrell_2016}.
In contrast, using the Plaintiff's plan with two VRA districts as a baseline would increase the pro-Republican bias of the enacted plans.
Freezing the three districts removes almost all variation in expected Republican seats in Alabama.

\end{document}